\documentclass[12pt]{article}

\usepackage[textwidth=450pt,textheight=650pt]{geometry}
\usepackage{amsmath}
\usepackage{amssymb}
\usepackage{bbold}

\makeatletter

\@addtoreset{equation}{section}
\makeatother
  
\begin{document}
\pagestyle{empty}
\begin{flushright}
 DCPT-08/47\\
August 2008
\end{flushright}
\vskip 7ex
\begin{center}
{\Large\bf Liouville Theory and Elliptic Genera\footnote{Contribution to the Proceedings of the conference `30 Years of Mathematical  Methods in High Energy Physics', Kyoto, March 2008, in honour of Prof. Tohru Eguchi's 60$^{th}$ birthday.}}\\ [7ex]
{\large Anne Taormina}\\[3ex]
{\em Department for Mathematical Sciences\\
Durham University,
South Road\,
Durham DH1 3LE,
United Kingdom}\\[3ex]
{\tt anne.taormina@durham.ac.uk}
\end{center}
\vskip 9ex

\begin{center}{\bf Abstract}
\end{center}
The structure and modular properties of ${\cal N}=4$ superconformal characters are reviewed and exploited, in an attempt to construct elliptic genera-like functions by decompactifying $K_3$.  The construction is tested against expressions obtained in the context of strings propagating in background ALE spaces of type $A_{N-1}$, using the underlying superconformal theory ${\cal N}=2$ minimal $\otimes$ $ {\cal N}=2$ Liouville. 

\vskip 2cm

\section{Introduction}
A better grasp of the conformal field theory underlying models of strings propagating on non-compact, singular space-times should provide interesting clues and contribute to their understanding. In this contribution to the Proceedings of Professor Eguchi's sixtieth birthday conference, we review our construction of new holomorphic modular invariants for asymptotically locally euclidean (ALE) spaces of type $A_{N-1}$\, \cite{EST06}\,. This construction is inspired by Witten's  notion of elliptic genus  of a compact Calabi-Yau manifold $M_d$, defined as the genus one partition function of the supersymmetric sigma model whose target space is $M_d$\,   \cite{Witten87}\,. Witten's elliptic genera are therefore expressible in terms of characters of  the underlying  superconformal algebras. A worldsheet description of strings in $A_{N-1}$ ALE backgrounds is believed to involve a pair of ${\cal N}=2$ superconformal theories (minimal and Liouville sectors)\, \cite{OV96}\, but, interestingly, some insights can be gained by decompactifying the $K_3$ manifold and exploit non-trivial properties of the representation theory and characters of the ${\cal N}=4$ superconformal algebra. \\

A central property of the relevant characters is their intricate behaviour under the modular group. Typically, characters come in two distinct species: `massless' characters
are discrete in number and encode topological information of the target space, while `massive' characters are, in particular, labelled by  a continuous, real parameter. Under the modular S-transformation, both species mix together, an observation made for the first time in 1987\,\cite{Eguchi87}\,, in the framework of the ${\cal N}=4$ superconformal algebra at central charge $c=6$. \\

In Section 2, we review the basic structure of 2-dimensional ${\cal N}=4$ superconformal characters and express their S-transformation in a way that highlights the role played by a remarkable integral due to Mordell \,\cite{Mordell33} in the construction of candidate conformal blocks in a theory with central charge $c=6$. For higher level $k\, (c=6k)$, we present some new material on generalisations of the Mordell integral. We also make a link between the expression for the S-transformation of level $k,\, {\cal N}=4$ Neveu-Schwarz (NS) characters presented in \eqref{St1} and \eqref{St2}, 
and the formula derived by Professor Eguchi in July 2005 and reproduced
 somewhere else\,\cite{EST08}\,. A few remarks relating our work to Zwegers' \, \cite{Zwegers02} 
conclude the section. Section 3 exploits the fact that ALE spaces are degenerate limits of the $K_3$ manifold. We start with a rewriting of the $K_3$ elliptic genus in terms of level $1$ ${\cal N}=4$ superconformal characters, and argue how to modify it to implement the relevant decompactification. This leads to two proposals for the definition of elliptic genera of $A_1$ and $A_{N-1}$ ALE spaces, whose consistency is tested against another approach in Section 4. There, we use  the description of $A_{N-1}$-type ALE spaces by the ${\mathbb Z}_N$ orbifolding of the tensor product of an ${\cal N}=2$ Liouville and an ${\cal N}=2$ minimal theory\, \cite{OV96}\,, and modify an earlier construction of the
corresponding elliptic genus\, \cite{ES04} according to the new insights on $\Gamma(2)$-completion obtained in the previous sections. Our two approaches at calculating the elliptic genus for $A_{N-1}$ ALE spaces result in two formulae which satisfy a highly non-trivial identity between theta functions. The latter was proven by Zagier\, \cite{Zagier}\,.

\section{Structure and modular properties of ${\cal N}=4$ superconformal characters}
The 2-dimensional ${\cal N}=4$ superconformal algebra contains four real supercharges arranged into two complex spinors of $SU(2)$, namely $(G^1, G^2)$ and $(\bar{G}^2, 
\bar{G}^1)$. Its commutation relations are, in terms of the Laurent modes,
\begin{eqnarray}
&&[L_m,L_n]=(m-n)L_{m+n}+\frac{1}{2}km(m^2-1)\delta_{m+n,0},\qquad m,n \in \mathbb Z,\nonumber\\
&&\{G^a_r,\,G^b_s\}=\{{\bar G}^a_r,\,{\bar G}^b_s\}=0,\qquad a,b = 1,2,\nonumber\\
&&\{G^a_r,\,{\bar G}^b_s\}=2\delta^{ab}L_{r+s}-2(r-s)\sigma^i_{ab}T^i_{r+s}+\frac{1}{2}k(4r^2-1)\delta_{r+s,0}\delta^{ab},\nonumber\\
&&[T^i_m,\,T^j_n]=i\epsilon^{ijk}T^k_{m+n}+\frac{1}{2}km\delta_{m+n,0}\delta^{ij},\nonumber\\
&&[T^i_m,\,G^a_r]=-\frac{1}{2}\sigma^i_{ab}G^b_{m+r},\qquad 
[T^i_m,\,{\bar G}^a_r]=\frac{1}{2}\sigma^{i*}_{ab}{\bar G}^b_{m+r},\nonumber\\
&&[L_m,\,G^a_r]=(\frac{1}{2}m-r)G^a_{m+r},\qquad [L_m,\,{\bar G}^a_r]=(\frac{1}{2}m-r){\bar G}^a_{m+r},\nonumber\\
&&[L_m,\,T^i_{n}]=-nT^i_{m+n}.\nonumber\\
\end{eqnarray}
where $T^i_m,\,i=1,2,3$ are the $\widehat{SU(2)}$, and $L_m$ the Virasoro, generators. 
The three matrices $\sigma^i$ are the Pauli matrices. The indices $r,s$ take integer values in the Ramond sector, and half-integer values in the Neveu-Schwarz sector. Consistency requires the central charge $c$ be quantized in units of 6. One writes  $c=6k$ with $k$  the level (taken to be a positive integer) of the $\widehat{SU(2)}$ affine subalgebra. The unitary highest weight state irreducible representations of this algebra have been extensively studied\, \cite{EguchiT1}\,. At given positive integer level $k$, they are labelled by the conformal weight $h$ and the isospin $\ell$ of the highest weight state $|\Omega>$. One has
\begin{equation}
L_0|\Omega> = h |\Omega>, \qquad T_0^3 |\Omega> = \ell |\Omega>,
\end{equation}
with $h \in \mathbb R$ satisfying the bounds
\begin{eqnarray}\label{bound}
&&h \ge k/4\qquad \qquad {\rm in\,the\,Ramond\, sector},\nonumber\\
&&h \ge \ell \qquad \qquad {\rm in\,the\,Neveu\,Schwarz\, sector},
\end{eqnarray}
and $\ell \in \frac{1}{2}\mathbb Z,\,\,0 \le \ell \le k/2$, as is well-known form the theory of $\widehat{SU(2)}$.
If the bound is saturated in \eqref{bound}, there exists a discrete number ($k+1$) of irreducible representations (called {\em massless}); if, on the other hand, the bound is not saturated, there exists a continuum of such representations (called {\em massive})\, \cite{Eguchi87}\,.

The characters associated with these highest weight irreducible representations are usually expressed as functions of two complex variables $\tau \in {\cal H}^+$ and $\mu \in \mathbb{C}$. Formally, one has
\begin{equation}
{\rm Ch}^{NS}_{h,k,\ell} (\tau, \mu) = Tr _{H}(\,e^{2i\pi \tau(L_0-c/24)} e^{2i\pi \mu T_0^3}\,).
\end{equation}

In the case of level $k=1$, the Neveu Schwarz (NS) massive characters can be written as,
\begin{multline}
{\rm Ch}^{NS}_{h,1,0}(\tau,\mu)=e^{2i\pi \tau (h-\frac{1}{4})}\,\prod_{n=1}^{\infty}\,\frac{(1+e^{2i\pi \mu}e^{2i\pi \tau (n-\frac{1}{2})})^2(1+e^{-2i\pi \mu} e^{2i\pi \tau (n-\frac{1}{2})})^2}{1-e^{2i\pi \tau n}}\\
=  e^{2i\pi \tau (h-1/8)}\,\frac{\vartheta_3 (\tau,\mu)^2}{\eta(\tau)^3}, \quad h >0,
\end{multline}
where $\eta(\tau)$ is the Dedekind function, and $\vartheta_3(\tau,\mu)$ one of Jacobi's theta functions.
This expression is consistent with the free field representation obtained in\, \cite{Matsuda}\,, which contains $SU(2)$ currents at level $k-1$, four free fermions and one free boson.
The structure of massless representations is more involved, due to the presence of fermionic null vectors. When $k=1$, there are two representations whose NS characters are
\begin{multline}
{\rm Ch}^{NS}_{\ell,1,\ell}(\tau,\mu)=e^{2i\pi \tau (\ell-1/8)}\frac{\vartheta_3(\tau, \mu)^2}{\eta(\tau)^3}\\
\times\,\,\prod_{n=1}^{\infty} \frac{1}{(1-e^{2i\pi \tau n})\,(1-e^{4i\pi \mu}e^{2i\pi \tau n})\,(1-e^{-4i\pi \mu}e^{2i\pi \tau (n-1)})}\\
\times \sum_{m \in \mathbb Z} e^{2i\pi \tau (2m^2+(2\ell+1)m)} \left\{ \frac{e^{2i\pi \mu (4m+2\ell)}}{(1+e^{2i\pi \mu}e^{2i\pi \tau(m+\frac{1}{2})})^2} -\frac{e^{-2i\pi \mu(4m+2\ell+2)}}{(1+e^{-2i \pi \mu}e^{2i\pi \tau (m+\frac{1}{2})})^2}
\right \},
\end{multline}
with $\ell =0$ and $\ell =1/2$.

The behaviour of the massless ${\cal N}=4$ characters under the modular transformation $S: \tau \rightarrow -\frac{1}{\tau}$ is interesting\, \cite{Eguchi87}\,. In the NS sector for instance, the transformation of the character corresponding to the highest isospin ($\ell =1/2$) can be written as
\begin{equation}\label{Stransform-1}
Ch_{\frac{1}{2},1,\frac{1}{2}}^{NS}(-\frac{1}{\tau},\frac{\mu}{\tau})=-e^{\frac{2i\pi\mu^2}{\tau}}{\rm Ch}^{NS}_{\frac{1}{2},1,\frac{1}{2}}(\tau,\mu)+ {\cal M},
\end{equation}
where the second term collects a continuum of ${\cal N}=4$ massive NS characters \break$Ch_{ \frac{\alpha^2}{2}+\frac{1}{8},1,0}^{NS},\, \alpha \in \mathbb R$, encoded in a remarkable integral. One has,
\begin{multline}\label{Stransform-2}
{\cal M} = e^{\frac{2i\pi\mu^2}{\tau}}\,\int d\alpha \,\,\frac{1}{2\cosh \pi \alpha}\,\, Ch_{ \frac{\alpha^2}{2}+\frac{1}{8},1,0}^{NS}(\tau,\mu)\\
= e^{\frac{2i\pi \mu^2}{\tau}}\frac{1}{\eta(\tau)}\,\int d\alpha\,\, \frac{e^{i\pi\alpha^2\tau}} {2\cosh \pi \alpha}\frac{\vartheta_3(\tau,\mu)^2}{\eta(\tau)^2}.
\end{multline}
The integral 
\begin{equation}
M(\tau)\equiv\frac{1}{\eta(\tau)}\,\int d\alpha\,\, \frac{e^{i\pi\alpha^2\tau}} {2\cosh \pi \alpha}
\end{equation}
was studied by Mordell in a different context in the 1930's, and was shown by him to be S-invariant. More specifically,  he rewrote\,\,  \cite{Mordell33} the integral in the explicitly S-invariant form $M(\tau)=h_3(\tau) + h_3(-1/\tau)$, with the function
\begin{equation}
h_3(\tau)=
\,\frac{e^{-i\pi \tau/4}}{\eta(\tau)\,\vartheta_3(\tau;0)}  \sum_{m \in \mathbb Z} \frac{e^{i\pi\tau m^2}}{1+e^{2i\pi\tau (m-1/2)}}
\end{equation}
directly related to a specialization of the level 1 Appell function  \cite{Appell} given by,
\begin{equation}\label{level1}
{\cal K}_{1}(\tau;\mu,\nu)=\sum_m\,\frac{e^{i\pi m^2  \tau+2i\pi m  \mu}}{1-e^{2i\pi(\mu+\nu+m\tau)}},\,\, \tau \in {\cal H}^+,\,\mu, \nu \in \mathbb C, \,\,\mu +\nu \notin \mathbb Z\tau +\mathbb Z.
\end{equation}
Indeed, one has
\begin{equation}
h_3(\tau)=\frac{e^{-i\pi\tau/4}}{\eta(\tau)\vartheta_3(\tau;0)}\,{\cal K}_{1}(\tau;0,-\frac{\tau}{2}-\frac{1}{2}).
\end{equation}

This allows for the construction of  a new function,
\begin{equation}\label{magic}
{\rm Ch}^{NS}_{\frac{1}{2},1,\frac{1}{2}}(\tau,\mu) - h_3(\tau)\,\frac{\vartheta_3(\tau,\mu)^2}{\eta(\tau)^2},
\end{equation}
whose behaviour under S is given by,
\begin{equation}
{\rm Ch}_{\frac{1}{2},1,\frac{1}{2}}^{NS}(-\frac{1}{\tau},\frac{\mu}{\tau})-
h_3(-\frac{1}{\tau})\,\frac{\vartheta_3(-\frac{1}{\tau},\frac{\mu}{\tau})^2}{\eta(-\frac{1}{\tau})^2}=-e^{\frac{2i\pi\mu^2}{\tau}}\,[\,{\rm Ch}^{NS}_{\frac{1}{2},1,\frac{1}{2}}(\tau,\mu) - h_3(\tau)\,\frac{\vartheta_3(\tau,\mu)^2}{\eta(\tau)^2}\,].
\end{equation}
The combination \eqref{magic} was recognized in\, \cite{Eguchi88}
as the following ratio of Jacobi theta functions,
\begin{equation}\label{rel1}
{\rm Ch}^{NS}_{\frac{1}{2},1,\frac{1}{2}}(\tau,\mu) - h_3(\tau)\,\frac{\vartheta_3(\tau,\mu)^2}{\eta(\tau)^2}=-\frac{\vartheta_1(\tau,\mu)^2}{\vartheta_3(\tau,0)^2},
\end{equation}
and it turns out that one also has
\begin{equation}\label{rel2}
{\rm Ch}^{NS}_{\frac{1}{2},1,\frac{1}{2}}(\tau,\mu) - h_4(\tau)\,\frac{\vartheta_3(\tau,\mu)^2}{\eta(\tau)^2}=\frac{\vartheta_2(\tau,\mu)^2}{\vartheta_4(\tau,0)^2}
\end{equation}
and
\begin{equation}\label{rel3}
{\rm Ch}^{NS}_{\frac{1}{2},1,\frac{1}{2}}(\tau,\mu) - h_2(\tau)\,\frac{\vartheta_3(\tau,\mu)^2}{\eta(\tau)^2}=-\frac{\vartheta_4(\tau,\mu)^2}{\vartheta_2(\tau,0)^2},
\end{equation}
where the functions $h_2(\tau)$ and $h_4(\tau)$ are related to two other specializations of the level 1 Appell function, namely
\begin{equation}
h_2(\tau)=\frac{1}{\eta(\tau)\vartheta_2(\tau,0)}\,{\cal K}_{1}(\tau;\frac{\tau}{2},-\frac{\tau}{2}-\frac{1}{2})
\end{equation}
and
\begin{equation}
h_4(\tau)=\frac{e^{-i\pi\tau/4}}{\eta(\tau)\vartheta_4(\tau,0)}\,{\cal K}_{1}(\tau;\frac{1}{2},-\frac{\tau}{2}-\frac{1}{2}).
\end{equation}
The three expressions \eqref{rel1},\eqref{rel2} and \eqref{rel3} can be seen as three rewritings of the massless character ${\rm Ch}^{NS}_{\frac{1}{2},1,\frac{1}{2}}(\tau,\mu)$. They provide a straightforward calculation of the Witten index of the massless representation, obtained by evaluating its NS character at $\mu=\frac{1}{2} (\tau+1)$. Since $\vartheta_3(\tau, \frac{1}{2} (\tau+1))=0$, one sees that the Witten index (i.e. the topological content of the massless representation) does not stem from the  term containing the function $h_3(\tau)$ (resp. $h_2(\tau)$ and $h_4(\tau)$). The three expressions  \eqref{rel1},\eqref{rel2} and \eqref{rel3} will play a crucial role in our discussion of elliptic genera later on. \\

Although we will not use ${\cal N}=4$ characters at level higher than 1 here, it is worth pointing out some properties of their S-transformation. The first derivation of the S-transformation known to me dates back to 2005 and is due to Professor Eguchi. It is based on the S-transformation
of a 3-parameter function ${\cal I}(p,a,b;\tau,\mu)$ \,\cite{Miki}\,, and was recently published \,\cite{EST08}\,. This elegant formula enables 
to express the S-transformation of superconformal characters in a way convenient to manipulate them. A slightly different starting point for the derivation is to express
the ${\cal N}=4$ characters in terms of higher level Appell functions, which were introduced in \, \cite{semi03} as generalizations of the level 1 Appell function \eqref{level1}. The level $p$ Appell function is defined as
\begin{equation}\label{Appell}
{\cal K}_{p}(\tau;\mu,\nu)=\sum_{m \in \mathbb Z}\,\frac{e^{i\pi m^2 p \tau+2i\pi m p \mu}}{1-e^{2i\pi(\mu+\nu+m\tau)}}, \qquad \tau \in {\cal H}^+,\,\, \mu, \nu \in \mathbb C, \,\,\mu+\nu \notin \mathbb Z \tau +\mathbb Z,
\end{equation}
with $p$  a positive integer. \\

The massless NS ${\cal N}=4$ character at level $k$ and isospin $\ell$ may be written as
\begin{multline}\label{charappell}
 {\rm Ch}^{NS}_{\ell, k,\ell}(\tau,\mu)=-ie^{-\frac{1}{2}(k+1)i\pi\tau}\frac{\vartheta_3^2(\tau,\mu)}{\eta^3(\tau)\vartheta_1(\tau, 2\mu)} \\
\times \sum_{n=0}^{2(k-2\ell+1)-1} (-1)^n
e^{2i\pi (\mu+\frac{\tau}{2})(2\ell+n+1)} {\cal K}_{2(k+1)}(\tau;\mu+\frac{2\ell+n+1}{2(k+1)}\tau,\frac{1}{2}+\frac{\tau}{2}-\frac{2\ell+n+1}{2(k+1)}\tau),
\end{multline}
and its S-transformation calculated from the knowledge of the modular properties of Appell functions\, \cite{semi03} which are stated in \eqref{StransformAppell}. The derivation of the S-transformation  of level $k$ ${\cal N}=4$ characters via Appell functions is technically involved and far less direct than the previous one,  and we do not  reproduce it here. However, it provides an expression for the S-transformation that is 
well suited for generalizing the Mordell integral, as we now discuss. 
The S-transformation via Appell functions is totally equivalent to that obtained in 2005 by Professor Eguchi (this, with hindsight, is not surprising since the level $p$ Appell function is closely related to the function ${\cal I}(p,a,b;\tau,\mu)$\footnote{For instance, ${\cal K}_p(\tau; \mu+\frac{1}{2}\tau, \frac{1}{2})=e^{-i\pi \tau p/4}e^{-i\pi \mu p}\,{\cal I}(p,0,0;\tau,\mu)$).}),
but whose connection to  the S-transformation of the level 1 characters as obtained in\, \cite{Eguchi87}\, is particularly straightforward, namely
\begin{equation} \label{St1}
{\rm Ch}^{NS}_{\ell, k, \ell} (-\frac{1}{\tau},\frac{\mu}{\tau})=(-1)^{2\ell}(k-2\ell+1) e^{\frac{2i\pi\mu^2k}{\tau}}\,{\rm Ch}^{NS}_{\frac{k}{2}, k,\frac{k}{2}}(\tau,\mu) + {\cal M}
\end{equation}
with
\begin{multline}\label{St2}
{\cal M}= \frac{1}{2}(-1)^{k-2\ell+1}e^{2i\pi\frac{\mu^2k}{\tau}}\,\sum_{a=1}^k \,(-1)^a\,\int_{\bf R}\,d\alpha\, {{\rm Ch}}^{NS}_{\frac{k+1}{4}\alpha^2+\frac{k+1}{4}(1+\frac{a-2}{k+1})^2,k,\frac{a-1}{2}}(\tau,\mu)\\
 \times \left\{ \sum_{n=0}^{k-2\ell} e^{\alpha n\pi +i\pi n} \frac{e^{\alpha \pi}\sin \frac{na}{k+1}\pi + \sin \frac{(n+1)a}{k+1}\pi}{\cosh \pi \alpha + \cos \frac{a}{k+1}\pi}\right\}.
\end{multline}
Indeed for $k=1, \ell =\frac{1}{2}$, \eqref{St1} and \eqref{St2} reduce to the level 1 result in \eqref{Stransform-2}. Using the following expression for massive characters \footnote{This definition differs from that of reference\, \cite{Eguchi87} by a Casimir factor, $e^{-\frac{k}{2}i\pi \tau}$.},
\begin{equation}
{\rm Ch}^{NS}_{h, k, \ell } (\tau,\mu)= e^{2i\pi\left\{h-\frac{(k+2\ell)^2}{4(k+1)}\right\}\tau}\,
\frac{\theta_3^2(\tau,\mu)}{\eta^3(\tau)}\chi_{\ell ,k-1}(\tau,\mu),\qquad h > \ell,
\end{equation}
where $\chi_{\ell,k}(\tau,\mu)$ are the $\widehat{SU(2)}$ characters
\begin{equation}
\chi_{\ell,k}(\tau,\mu)=\frac{\theta_{2\ell+1,k+2}(\tau,2\mu)-\theta_{-2\ell-1,k+2}(\tau,2\mu)}{\theta_{1,2}(\tau,2\mu)-\theta_{-1,2}(\tau,2\mu)},
\end{equation}
one obtains
\begin{multline}\label{massive}
{\cal M}= \frac{1}{2}(-1)^{k-2\ell+1}e^{2i\pi\frac{\mu^2k}{\tau}}\,\sum_{a=1}^k\,(-1)^a\,\frac{1}{\eta(\tau)} \int_{\bf R}\,d\alpha\,e^{i\pi \frac{k+1}{2}\alpha^2\tau}\,\\
\times \left\{ \sum_{n=0}^{k-2\ell} e^{\alpha n\pi +i\pi n} \frac{e^{\alpha \pi}\sin \frac{na}{k+1}\pi + \sin \frac{(n+1)a}{k+1}\pi}{\cosh \pi \alpha + \cos \frac{a}{k+1}\pi}\right\}\, \chi_{\frac{1}{2}(a-1),k-1}(\tau,\mu)\,\frac{\vartheta_3(\tau,\mu)^2}{\eta(\tau)^2}.
\end{multline}

We show in the Appendix how to relate this expression to that appearing in \, \cite{EST08}\,. It is interesting to note that \eqref{massive} provides  generalizations of the Mordell integral encountered above.
Indeed, taking 
 $\ell = \frac{k}{2}$ in the above expression yields the relation,

\begin{multline} \label{k2l}
{\rm Ch}^{NS}_{\frac{k}{2}, k, \frac{k}{2}} (-\frac{1}{\tau},\frac{\mu}{\tau})=(-1)^{k} e^{\frac{2i\pi\mu^2k}{\tau}}\,{\rm Ch}^{NS}_{\frac{k}{2}, k,\frac{k}{2}}(\tau,\mu) \\
-\frac{1}{\eta(\tau)} e^{\frac{2i\pi\mu^2k}{\tau}} \sum_{a=1}^k (-1)^a\int d\alpha\,e^{i\pi \frac{k+1}{2}\alpha^2\tau}\,\\
\times \frac{\sin\,\frac{a\pi}{k+1}}{2\cosh\,\pi\alpha+2 \cos\,\frac{a\pi}{k+1}}\, \chi_{\frac{1}{2}(a-1),k-1}(\tau,\mu)\,\frac{\vartheta_3(\tau,\mu)^2}{\eta(\tau)^2},
\end{multline}
which, for $k=1$, reproduces the results quoted in \eqref{Stransform-1} and \eqref{Stransform-2}, and obtained in 1988 for $c=6$\,\cite{Eguchi87}\,. The expression \eqref{k2l} contains $k$ generalized Mordell integrals
\begin{equation}
  M_{a,k}(\tau)=\frac{1}{\eta(\tau)}\,\int \,d\alpha\,e^{i\pi \frac{k+1}{2}\alpha^2\tau}\,\frac{\sin\,\frac{a\pi}{k+1}}{2\cosh\,\pi\alpha+2 \cos\,\frac{a\pi}{k+1}},\qquad a=1,..,k,
  \end{equation}
whose relation to their  S-transformation is \footnote{The proof, obtained in collaboration with T. Eguchi (March 2008), relies on standard techniques of Fourier transforms.}
 \begin{equation}
   \sqrt{\frac{2}{k+1}}\,\sum_{b=1}^k (-1)^{a+b}\,\sin\,\frac{ab\pi}{k+1}\,M_{b,k}(-\frac{1}{\tau})= (-1)^{k+1}\,M_{a,k}(\tau).
   \end{equation}
The consequences of this property in the context of ${\cal N}=4$ superconformal field theory are under investigation at present.\\

We end this section with a further remark. 
The level 1 Appell function (and consequently the function $h_3(\tau)$) are related to a function $\mu$ studied by Lerch\, \cite{Lerch86}\,, 
\begin{equation}\label{Lerch}
\mu(\tau; u,v)=-\frac{e^{i\pi u}}{\vartheta_1(\tau,v)}\,{\cal K}_{1}(\tau;v+\frac{\tau+1}{2},u-v-\frac{\tau+1}{2}),
\end{equation}
whose important properties are rederived in Zwegers' thesis \, \cite{Zwegers02}\,. 
Specifically, one has,
\begin{equation}
h_3(\tau)=\frac{i}{ \eta(\tau)}\,\mu(\tau;-\frac{\tau+1}{2},-\frac{\tau+1}{2}).
\end{equation}
Zwegers proceeds to construct a completion of  the Lerch sum $\mu(\tau;u,v)$ by a real analytic function $R(\tau;u-v)$ so that the resulting function $\tilde{\mu}(\tau;u,v)$ transforms under $S$ exactly like our combination \eqref{magic}. 
\section{$K_3$ elliptic genus and ${\cal N}=4$ characters}

The aim of this section is to express the elliptic genus of the compact $K_3$ manifold in terms of level 1 ${\cal N}=4$ superconformal characters, and use the information encoded in this rewriting to propose a mathematical expression, which could be defined as the elliptic genus of an $A_1$ ALE space.\\

The $K_3$ elliptic genus may be derived from an orbifold calculation on $T^4/{\mathbb Z}_2$ \cite{Eguchi88} and written as,
\begin{equation}
Z_{K3}(\tau,\mu)=8 \left[ \left( \frac{\vartheta_3(\tau,\mu)}{\vartheta_3(\tau,0)}\right)^2+ \left( \frac{\vartheta_4(\tau,\mu)}{\vartheta_4(\tau,0)}\right)^2+ \left( \frac{\vartheta_2(\tau,\mu)}{\vartheta_2(\tau,0)}\right)^2\right].
\end{equation}
In order to make our point, we spectral flow this expression to the NS sector
\begin{equation}\label{spectralflow}
e^{2i\pi(\frac{\tau}{4}-\mu)}\,Z_{K3}(\tau,\mu-\frac{1}{2}(\tau+1))=8\left [ -\left ( \frac{\vartheta_1(\tau,\mu)}{\vartheta_3(\tau,0)}\right)^2+\left ( \frac{\vartheta_2(\tau,\mu)}{\vartheta_4(\tau,0)}\right)^2-\left ( \frac{\vartheta_4(\tau,\mu)}{\vartheta_2(\tau,0)}\right)^2\right],
\end{equation}
and use the relations \eqref{rel1}, \eqref{rel2} and \eqref{rel3} obtained in the previous section to rewrite \eqref{spectralflow} as,
\begin{equation}
e^{2i\pi(\frac{\tau}{4}-\mu)}\,Z_{K3}(\tau,\mu-\frac{1}{2}(\tau+1))=24{\rm Ch}_{\frac{1}{2},1,\frac{1}{2}}^{NS}(\tau,\mu)-8\,\eta(\tau)\,\sum_{i=2,3,4}h_i(\tau)\frac{\theta_3(\tau,\mu)^2}{\eta(\tau)^3}.
\end{equation}
By isolating a contribution $2e^{-\frac{i\pi\tau}{4}}$ in $8\eta(\tau)\,h_2(\tau)$, one may write
\begin{equation}
8\eta(\tau)\sum_{i=2,3,4}h_i(\tau)=e^{-\frac{i\pi\tau}{4}}[2-\sum_{n=1}^{\infty} a_n\, e^{2i\pi n\tau}]
\end{equation}
where all coefficients $a_n$ are positive integers \, \cite{Wendland00}\,. Furthermore, exploiting a well-known property of ${\cal N}=4$ characters\, \cite{EguchiT1}\,, namely
\begin{equation}\label{massless-massive}
2{\rm Ch}_{0,1,0}^{NS}(\tau,\mu)+4\,{\rm Ch}_{\frac{1}{2},1,\frac{1}{2}}^{NS}(\tau,\mu)=2e^{-\frac{i\pi \tau}{4}}\frac{\theta_3(\tau,\mu)^2}{\eta(\tau)^3},
\end{equation}
the spectral-flowed elliptic genus for $K_3$ becomes
\begin{multline}\label{ellipticgenus-2}
e^{2i\pi(\frac{\tau}{4}-\mu)}\,Z_{K3}(\tau,\mu-\frac{1}{2}(\tau+1))=\\
20 {\rm Ch}^{NS}_{\frac{1}{2},1,\frac{1}{2}}(\tau,\mu)-2{\rm Ch}^{NS}_{0,1,0}(\tau,\mu)+\sum_{n=1}^{\infty}a_n\,e^{2i\pi \tau (n-1/8)} \frac{\theta_3(\tau,\mu)^2}{\eta(\tau)^3}.
\end{multline}
In order to proceed towards our goal of providing a candidate elliptic genus for $A_1$ ALE spaces, we exploit the property of $K_3$ manifolds to be decomposable into a sum of 16 $A_1$ ALE spaces \,\cite{Page78}\,. Such decompactification is achieved by decoupling gravity, and in the language of ${\cal N}=4$ superconformal theory, amounts to
removing  from the expression \eqref{ellipticgenus-2} the contribution stemming from the massless NS representation with character
${\rm Ch}_{0,1,0}^{NS}(\tau,\mu)$ (isospin $\ell=0$, which contains the graviton).\\

Since the appearance of this representation in the spectral-flowed elliptic genus is due to the rewriting (using \eqref{massless-massive}) of the term  $2e^{-\frac{i\pi\tau}{4}} \frac{\theta_3(\tau,\mu)^2}{\eta(\tau)^3}$ originating from the function 
$8\eta(\tau)\,h_2(\tau)\frac{\theta_3(\tau,\mu)^2}{\eta(\tau)^3}$, we propose, given \eqref{rel3},  to drop the contribution $\left( \frac{\vartheta_4(\tau,\mu)}{\vartheta_2(\tau,0)}\right)^2$ from \eqref{spectralflow} and obtain
\begin{equation}\label{decomp}
e^{2i\pi(\frac{\tau}{4}-\mu)}\,Z_{K_3}^{decomp}(\tau,\mu-\frac{1}{2}(\tau+1))=8\left [ -\left ( \frac{\vartheta_1(\tau,\mu)}{\vartheta_3(\tau,0)}\right)^2+\left ( \frac{\vartheta_2(\tau,\mu)}{\vartheta_4(\tau,0)}\right)^2\right].
\end{equation}
Spectral-flowing back to the $\tilde R$ sector, we  arrive at the following expression for the elliptic genus of the decompactified $K_3$ manifold,
\begin{equation}
Z_{K_3}^{decomp}(\tau,\mu)=8\left[ \left(\frac{\theta_3(\tau,\mu)}{\theta_3(\tau,0)}\right)^2+\left(\frac{\theta_4(\tau,\mu)}{\theta_4(\tau,0)}\right)^2\right].
\end{equation}
 Two proposals follow from the chain of arguments above.\\

{\bf \em Proposal 1}: The elliptic genus for the $A_1$ ALE space is
\begin{equation}
Z_{A_1}(\tau,\mu)=\frac{1}{2}\left[ \left(\frac{\vartheta_3(\tau,\mu)}{\vartheta_3(\tau,0)}\right)^2+\left(\frac{\vartheta_4(\tau,\mu)}{\vartheta_4(\tau,0)}\right)^2\right].
\end{equation}

{\bf \em Proposal 2}: The elliptic genus for the $A_{N-1}$ ALE space is
\begin{equation}\label{proposal2}
Z_{A_{N-1}}(\tau,\mu)=\frac{N-1}{2}\left[ \left(\frac{\vartheta_3(\tau,\mu)}{\vartheta_3(\tau,0)}\right)^2+\left(\frac{\vartheta_4(\tau,\mu)}{\vartheta_4(\tau,0)}\right)^2\right].
\end{equation}

In the next section, we test the consistency of these proposals against another approach at calculating elliptic genera for $A_{N-1}$ ALE spaces.\\

An off-shoot of this construction is to propose conformal blocks for non-compact conformal field theories. Indeed  when  \eqref{decomp} is rewritten in terms of NS characters as,
\begin{equation}
e^{2i\pi(\frac{\tau}{4}-\mu)}\,Z_{A_1}(\tau,\mu-\frac{1}{2}(\tau+1))=
 {\rm Ch}^{NS}_{\frac{1}{2},1,\frac{1}{2}}(\tau,\mu)-\frac{1}{2}\eta(\tau)\left[h_3(\tau)+h_4(\tau)\right]
\frac{\vartheta_3(\tau,\mu)^2}{\eta(\tau)^3},
\end{equation}
one can show\,\cite{Eguchi08} that 
\begin{equation}
\frac{1}{2}\eta(\tau)\sum_{i=3,4}h_i(\tau)=-\sum_{n=1}b_nq^{n-1/8},
\end{equation}
with all coefficients $b_n$ being positive. It is therefore natural to suggest that the  following `$\Gamma(2)$-invariant\footnote{$\Gamma(2)$ is the principal congruence subgroup of level 2 of $SL(2,\mathbb Z)$.} completion of the NS ${\cal N}=4$ massless character 
for $\ell=1/2$',
\begin{multline}\label{confblock}
\left[ {\rm Ch}^{NS}_{\frac{1}{2},1,\frac{1}{2}}(\tau,\mu) \right]_{\Gamma(2)-{\rm inv}}=
{\rm Ch}^{NS}_{\frac{1}{2},1,\frac{1}{2}}(\tau,\mu)-\frac{1}{2}\eta(\tau)\left[ h_3(\tau)+h_4(\tau)\right]\frac{\vartheta_3(\tau,\mu)^2}{\eta(\tau)^3}\\
\equiv \frac{1}{2}\left [ -\left ( \frac{\vartheta_1(\tau,\mu)}{\vartheta_3(\tau,0)}\right)^2+\left ( \frac{\vartheta_2(\tau,\mu)}{\vartheta_4(\tau,0)}\right)^2\right]
\end{multline}
be taken as a conformal block of the underlying theory. Note that the $\Gamma(2)$-invariant completion selects the topological content of massless representations, as is
clear from the comments appearing beneath \eqref{rel3}. We also stress that mathematically, the $\Gamma(2)$-invariant completion of $ {\rm Ch}^{NS}_{\frac{1}{2},1,\frac{1}{2}}(\tau,\mu)$ is not unique. For instance, one could envisage including a term $\left ( \vartheta_4(\tau,\mu))/\vartheta_2(\tau,0)\right)^2$ in \eqref{confblock}. However, in the present physical context, one can lift this ambiguity by studying the equivalent of expression \eqref{confblock}  in the $\tilde R$ sector, namely
\begin{equation}
\left[ {\rm Ch}^{\tilde R}_{\frac{1}{4},1,0}(\tau,\mu) \right]_{\Gamma(2)-{\rm inv}}=
 \frac{1}{2}\left [ \left ( \frac{\vartheta_3(\tau,\mu)}{\vartheta_3(\tau,0)}\right)^2+\left ( \frac{\vartheta_4(\tau,\mu)}{\vartheta_4(\tau,0)}\right)^2\right].
 \end{equation}
The right-hand side corresponds to a GSO projection ensuring that  the $q$-expansion is integer-powered, as required in the Ramond sector. Adding a term $(\vartheta_2(\tau,\mu)/\vartheta_2(\tau,0))^2$ (equivalent to a term $\left ( \vartheta_4(\tau,\mu))/\vartheta_2(\tau,0)\right)^2$ in the NS sector), would in particular introduce a massive representation whose conformal weight would be below threshold ($h=0 < 1/4$), as such a term is associated with $h_2(\tau)$ through the relation
\begin{equation}
{\rm Ch}^{\tilde{R}}_{\frac{1}{4},1,0}(\tau,\mu)-h_2(\tau)\frac{\vartheta_1(\tau,\mu)^2}{\eta(\tau)^2}=\frac{\vartheta_2(\tau,\mu)^2}{\vartheta_2(\tau,0)^2}.
\end{equation}
\section{Elliptic genera and tensor products of Liouville and minimal ${\cal N}=2$ models}

A-type ALE spaces, which are obtained by blowing up $A_{N-1}$ singularities, may be described by the ${\mathbb Z}_N$ orbifolding of the tensor product of the ${\cal N}=2$ Liouville theory $L_N$ with central charge $\hat{c}_L=1+\frac{2}{N}$ and the  ${\cal N}=2$ minimal model $M_k$ with central charge $\hat{c}_M=1-\frac{2}{N}, N=k+2$\, \cite{OV96}\,.\\

We can therefore attempt to compute the elliptic genus of A-type ALE spaces by pairing contributions from the ${\cal N}=2$ minimal and Liouville theories. \\

Recall that the elliptic genus $Z(\tau,\mu)$ is defined by taking the sum over all states in the left-moving sector of the theory, while the right-moving sector is fixed at the Ramond ground states\, \cite{Witten87}\,. Namely,
\begin{equation} \label{ellgenus}
Z(\tau; \mu)={\rm Tr}_{R \otimes R}(-1)^{F_L+F_R}e^{2i\pi \mu T_0^{3L}}\,e^{2i\pi \tau (L_0-\frac{c}{24})}e^{-2i\pi \bar{\tau}(\bar{L}_0-\frac{c}{24})},
\end{equation}
with $T_0^{3L}$ denoting the $U(1)_R$ charge in the left-moving sector and the trace being taken in the Ramond-Ramond sector.  \\

The contribution from the minimal sector is straightforward. If we follow the prescription
\eqref{ellgenus}, it reads,
\begin{equation}
Z_{{\rm minimal}}(\tau, \mu)=\sum_{\ell =0}^{N-2}{ \rm Ch}^{\tilde {R}}_{\ell, \ell+1}(\tau,\mu),
\end{equation}
where ${ \rm Ch}^{\tilde {R}}_{\ell, \ell+1}(\tau,\mu)$ denote the ${\cal N}=2$ minimal model characters associated to the Ramond ground state\, \cite{Yang}\,, and the $\tilde{R}$ symbol refers to the Ramond sector with $(-1)^F$ insertion, $F$ being the fermionic number. There is however a shortcut to this procedure, based on the fact the Landau-Ginzburg theory described by the superpotential
\begin{equation}
W=g\,(X^{k+2}+Y^2+Z^2)
\end{equation}
acquires scale invariance in the infrared limit, and reproduces the ${\cal N}=2$ minimal theory with $c_M=1-\frac{2}{k+2}$\,\, \cite{WittenLG}\,. As the coupling constant $g$ tends to 0, the Landau-Ginzburg theory becomes the theory of a free chiral field $X$ with $U(1)_R$ charge $\frac{1}{N}$ (recall $N=k+2$).  Thus the elliptic genus contribution includes that of a free boson of charge $\frac{1}{N}$ and of a free fermion of charge $\frac{1}{N}-1$, and one actually has\,\cite{WittenLG}\,,
\begin{equation}\label{LG}
Z_{LG}(\tau,\mu)=\frac{\vartheta_1(\tau,(1-\frac{1}{N})\mu)}{\vartheta_1(\tau,\frac{1}{N}\mu)}=Z_{{\rm minimal}}(\tau, \mu),\qquad N=k+2.
\end{equation}

The contribution from the Liouville sector is more involved. It is given by the sum of $N$ extended discrete characters $\chi_{{\rm discrete}}^{\tilde R}(s,s-1;\tau,\mu), s=1,..,N$, which can be expressed in terms of the Appell function at level $2N$ (see \eqref{Appell}) in the following way\,
\cite{ES04b}\,,
\begin{equation}\label{Liouville}
Z_{{\rm Liouville}}(\tau,\mu)=\sum_{s=1}^N\chi_{{\rm dis}}^{\tilde{R}}(s,s-1;\tau,\mu)
= {\cal K}_{2N}(\tau;\frac{\mu}{N},0)\,\frac{\vartheta_1(\tau,\mu)}{\eta(\tau)^3}.
\end{equation}
A striking contrast between the contribution from the minimal  and the Liouville sectors is that the latter does not enjoy `good' modular properties, as a direct consequence of the S-transformation law of the level $p$ Appell function\, \cite{semi03}\,,
\begin{multline}\label{StransformAppell}
{\cal K}_p(-\frac{1}{\tau};\frac{\mu}{\tau},\frac{\nu}{\tau})=\tau\,e^{i\pi p \frac{\mu^2-\nu^2}{\tau}}\,{\cal K}_p(\tau;\mu,\nu)\\
+\tau\, \sum_{a=0}^{p-1}\,e^{i\pi \frac{p}{\tau}(\mu+\frac{a}{p}\tau)^2}\,\Phi(p\tau,p\nu-a\tau)\,\vartheta(p\tau,p\mu+a\tau)
\end{multline}
where 
\begin{equation}
\Phi(\tau,\mu)=-\frac{i}{2\sqrt{-i\tau}}-\frac{1}{2}\,\int_{\mathbb R}\,dx\,e^{-\pi x^2} \frac{\sinh(\pi x\sqrt{-i\tau}(1+2\frac{\mu}{\tau})}{\sinh (\pi x\sqrt{-i\tau})}.
\end{equation}
It is therefore not surprising that a `naive' construction for the elliptic genus of $A_{N-1}$
spaces, which involves orbifolding the contributions from the minimal and Liouville sectors according to\, \cite{ES04}\,,
\begin{multline}\label{orbi}
Z_{ALE(A_{N-1})}(\tau,\mu)=\frac{1}{N}\sum_{a,b \in {\mathbb Z}_N}\,e^{2i\pi \tau a^2}e^{4i\pi a\mu}Z_{{\rm minimal}}(\tau,\mu+a\tau+b)\\
\times \,Z_{{\rm Liouville}}(\tau,\mu+a\tau+b)
\end{multline}
yields a formula which does  not have a nice behaviour under the modular group.\\

In fact, the elliptic genus is associated with a conformal field theory defined on the torus, and hence it must be invariant under $SL(2,{\mathbb Z})$ or under one of its subgroups. Since we are dealing with a superconformal field theory, it seems natural  to demand invariance under $\Gamma(2)$, which leaves  the spin structures fixed. Therefore, we proceed to construct an elliptic genus invariant under $\Gamma(2)$, which is generated by $T^2$ and $ST^2S^{-1}$.\\

To do so, we propose to replace the level $2N$ Appell function appearing in $Z_{{\rm Liouville}}(\tau,z+a\tau+b)$  by a $\Gamma(2)$-invariant completion. It turns out that the desired completion, whose derivation is presented elsewhere \,\cite{EST08}\,, is given by
\begin{equation}\label{modifiedAppell}
\left[{\cal K}_{2N}(\tau,\mu)\right]_{\Gamma(2)-\rm inv} \equiv \frac{1}{4} \frac{i\eta(\tau)^3\vartheta_1(\tau,2\mu)}{\vartheta_1(\tau,\mu)^2}\left[\, \left( \frac{\vartheta_3(\tau,\mu)}{\vartheta_3(\tau,0)}\right)^{2(N-1)}+\left( \frac{\vartheta_4(\tau,\mu)}{\vartheta_4(\tau,0)}\right)^{2(N-1)}\,\right].
\end{equation}

Inserting \eqref{modifiedAppell} in \eqref{orbi}, using the explicit formulas \eqref{LG} and \eqref{Liouville}, we arrive at a third proposal, namely,\\

{\bf \em Proposal 3}: The elliptic genus for $A_{N-1}$ ALE spaces is given by
 \begin{multline}\label{proposal3}
Z_{ALE(A_{N-1})}(\tau,\mu)=
\frac{1}{4N}\sum_{a,b=1}^Ne^{i\pi\tau{a^2}}e^{2i\pi a\mu} (-1)^{a+b}\frac{\displaystyle\vartheta_1(\tau,\frac{N-1}{N}\mu_{a,b})\,\vartheta_1(\tau,\frac{2}{N}\mu_{a,b})\,\vartheta_1(\tau,\mu)}{\displaystyle\vartheta_1(\tau,\frac{1}{N}\mu_{a,b})^3}\\
\times \left[ 
\left(\frac{\displaystyle\theta_3(\tau,\frac{1}{N}\mu_{a,b})}{\displaystyle\vartheta_3(\tau,0)}\right)^{2(N-1)}+\left(\frac{\displaystyle\vartheta_4(\tau,\frac{1}{N}\mu_{a,b})}{\displaystyle\vartheta_4(\tau,0)}\right)^{2(N-1)}\right],
\end{multline}
where $\mu_{a,b}\equiv \mu+a\tau+b$.\\

Remarkably, the expressions \eqref{proposal2} and  \eqref{proposal3} are equal. It is easy to check the case $N=2$ using addition theorems of theta functions, but a mathematical proof for higher values of $N$ is much harder. Don Zagier found an elegant proof which uses residue integrals\, \cite{Zagier}\,, and which was reproduced in an appendix of a previous paper\, \cite{EST06}\,. This provides us with a robust consistency check of the arguments we used to arrive at
a construction of elliptic genera for certain non-compact Calabi-Yau manifolds.  
\section{Conclusion}
A string theory propagating on a non-compact Calabi-Yau manifold is described by  a 2-dimensional conformal field theory possessing two types of representations, called massless and massive. Massless representations  encode topological information of the target manifold, and their number is discrete. Massive representations are believed to be related to deformations of the moduli, and there exists a continuous infinity of them. The corresponding massless and massive characters transform in an interesting way under the modular S-transformation, schematically represented as\\
\begin{eqnarray}
{\rm massless\, (discrete)}\qquad&\stackrel{S}{\rightarrow}&\qquad \sum\,\,{\rm massless\,(discrete)} + \int\,{\rm massive\,(continuous)}\nonumber\\
{\rm massive\,(continuous)}\qquad&\stackrel{S}{\rightarrow} &\qquad \int \,{\rm massive\,(continuous)}\nonumber
\end{eqnarray}
The mathematical structure of such transformations is traceable to the modular behaviour of Appell functions\,\cite{Appell, Polischuk,semi03}\,, and our attempt at constructing conformal blocks is similar in spirit to the work done by Zwegers in his PhD thesis, where a Lerch sum\, \cite{Lerch86}\, (related to the level 1 Appell function through \eqref{Lerch}) is completed by a real-analytic function to produce a function which transforms as a two-variable Jacobi form of weight 
$\frac{1}{2}$ and index $\left( \begin{array}{cc} -1&1\\1&-1 \end{array} \right)$.
This completion technique has triggered a  breakthrough in the theory of mock theta functions \,\cite{Zwegers02, Ono} and a spurt of activity in the Number Theory community in recent years. On the physical interpretation of our $\Gamma(2)$-completions, much remains to be understood. In particular, the geometric significance of our proposed elliptic genera for $A_{N-1}$ ALE spaces is far from elucidated.\\

\section*{Acknowledgements}
I would like to thank Prof. Tohru Eguchi for sharing his insights into Theoretical Physics with me since 1986, and for a long-lasting collaboration on superconformal characters.
I also wish to thank the Organizing Committee for setting up such a stimulating conference to celebrate Prof. Eguchi's 60th birthday in Kyoto last spring, and the Yukawa
Institute for their warm hospitality.

\appendix
\section{} 
We show how to relate the S-transformation of ${\cal N}=4$ NS characters at arbitrary positive integer level $k$ stated in \eqref{massive} to the derivation presented in Appendix C of \, \cite{EST08}\,. From the information given in the formula (C.16) therein, one can write the S-transformation as
\begin{multline}
{\rm  Ch}_{\ell,k,\ell}^{NS}(-1/\tau,\mu/\tau)=(-1)^{2\ell}(k-2\ell+1){\rm Ch}_{k/2,k,k/2}^{NS}(\tau,\mu)\,+\,
ie^{\frac{2i\pi k \mu^2}{\tau}}\frac{\vartheta_3(\tau,\mu)^2}{\eta(\tau)^3}\\
\times
\sum_{j=0}^{2k+1}\frac{(-1)^j}{ \sqrt{2(k+1)}}\, \chi_{\frac{1}{2}(j-1),k-1}(\tau,\mu)\,\int \,dp \,q^{\frac{p^2}{2}}
\frac{  e^{-2\pi(2\ell-k)\left(   \frac{p}{\sqrt{2(k+1)}}+    i\frac{j}{ 2(k+1)} \right)     }       }
{ \left[  1+e^{-2\pi\left(\frac{p}{ \sqrt{2(k+1)}}+i\frac{j}{ 2(k+1)}\right)  } \right]^2}.
 \end{multline}
\\
The massive contribution above can also be rewritten as a sum over $j=1,.., k$ as follows:
\begin{multline}\label{massiveMiki}
{\cal M}'= ie^{\frac{2i\pi k \mu^2}{\tau}}\sum_{j=1}^{k}\frac{(-1)^j}{ \sqrt{2(k+1)}}\, \chi_{\frac{1}{2}(j-1),k-1}(\tau,\mu)\,\frac{\vartheta_3(\tau,\mu)^2}{\eta(\tau)^3}\,\int dp\,q^{\frac{p^2}{2}}\\
\times \left\{
\frac{ e^{-2\pi(2\ell-k)  \left(\frac{p}{ \sqrt{2(k+1)}}+i\frac{j}{ 2(k+1)}\right)} }
{ \left[ 1+e^{-2\pi \left(\frac{p}{ \sqrt{2(k+1)}}+i\frac{j}{ 2(k+1)}\right)}\right]^2}
-
\frac{e^{-2\pi(2\ell-k) \left(\frac{p}{ \sqrt{2(k+1)}}-i\frac{j}{ 2(k+1)}\right)} }
{ \left[1+e^{-2\pi \left(\frac{p}{ \sqrt{2(k+1)}} -i\frac{j}{2(k+1)}\right)}\right]^2}
\right\}.
\end{multline} 
Introducing the notation,
\begin{equation}\label{notations}
X_r\equiv e^{-2\pi\left({p\over \sqrt{2(k+1)}}+i{r\over 2(k+1)}\right)},
\end{equation}
\eqref{massiveMiki} takes the form,
\begin{multline}
{\cal {M}}'=\frac{i}{\sqrt{2(k+1)}}\,e^{2i\pi \frac{\mu^2}{\tau}k}\, 
\sum_{j=1}^k (-1)^j\, \chi_{\frac{1}{2}(j-1),k-1}(\tau,\mu)\,\frac{\vartheta_3(\tau,\mu)^2}{\eta(\tau)^3}\\
\times
\int \, dp\,q^{\frac{p^2}{2}}\,\frac{X_j^{2\ell-k}(1+X_{-j})^2-X_{-j}^{2\ell-k}(1+X_{j})^2}{(1+X_j)^2(1+X_{-j})^2}.
\end{multline}
On the other hand, using \eqref{notations} and the change of variable $\alpha=\frac{2}{\sqrt{2(k+1)}}p$ in \eqref{massive}, we obtain
\begin{multline}
{\cal {M}}=\frac{-i}{\sqrt{2(k+1)}}\,
(-1)^{k-2\ell+1}\,\,e^{2i\pi \frac{\mu^2}{\tau}k}\, 
\sum_{a=1}^k (-1)^a\, \chi_{\frac{1}{2}(a-1),k-1}(\tau,\mu)\,\frac{\vartheta_3(\tau,\mu)^2}{\eta(\tau)^3}\\
\times
\int \, dp\,q^{\frac{p^2}{2}}\,\left\{ \sum_{n=0}^{k-2\ell}\,(-1)^n
\frac{X_a^{-n}-X_{-a}^{-n}+X_a^{-n}X_{-a}-X_{-a}^{-n}X_a}{(1+X_a)(1+X_{-a})}
\right\}.
\end{multline}
Since $\sum_{n=0}^{k-2\ell}(-1)^nX^n=\displaystyle{\frac{1+(-1)^{k-2\ell}X^{k-2\ell+1}}{1+X}}$, a short succession of elementary manipulations yields
\begin{multline}
\sum_{n=0}^{k-2\ell}\,(-1)^n
\frac{X_a^{-n}-X_{-a}^{-n}+X_a^{-n}X_{-a}-X_{-a}^{-n}X_a}{(1+X_a)(1+X_{-a})}=\\
\frac{(-1)^{k-2\ell}X_a^{2\ell-k}(1+X_{-a})^2-(-1)^{k-2\ell}X_{-a}^{2\ell-k}(1+X_{a})^2
{+(X_a-X_{-a})(1-X_aX_{-a})}}{(1+X_a)^2(1+X_{-a})^2}.
\end{multline}
The massive contributions ${\cal M}$ and ${\cal M}'$ coincide as the terms
\begin{equation}
(X_a-X_{-a})(1-X_aX_{-a})/(1+X_a)^2(1+X_{-a})^2
\end{equation}
are odd under $p \rightarrow -p$ and do not survive the integration over momentum  \break $p \in ]-\infty,+\infty[$  with weight $q^{\frac{p^2}{2}}$.
%

\end{document}